\documentclass[11pt]{article}
\usepackage{jheppubmod}
\pdfoutput=1
\usepackage{comment}
\usepackage{framed}
\usepackage{psfrag}
\usepackage{array}
\usepackage{bm, bbm, bbold}
\usepackage{amsmath,amsfonts,amssymb, mathrsfs, amsthm}
\usepackage{braket}
\usepackage{graphicx}
\usepackage[labelsep=quad]{subcaption}
\usepackage{color,soul}
\usepackage[punctsep]{collref}
\collectsep[]{;}
\usepackage{epsfig}
\usepackage{wrapfig}
\usepackage{tikz}
\usetikzlibrary{arrows,plotmarks,positioning,calc}
\usepackage{environ}

\NewEnviron{eqn}{
\begin{align}
\begin{split}
  \BODY
\end{split}
\end{align}
}
\NewEnviron{eqn*}{
\begin{align*}
\begin{split}
  \BODY
\end{split}
\end{align*}
}
\newcommand{\be}{\begin{equation}}
\newcommand{\ee}{\end{equation}}
\newcommand{\bes}{\begin{equation*}}
\newcommand{\ees}{\end{equation*}}

\newcommand{\tf}{\mbox{\footnotesize tf}}
\newcommand{\intsp}{\int_{\mathcal I^+}}
\newcommand{\stress}{\mathcal{T}}
\newcommand{\tila}{\tilde{\mrm a}}
\newcommand{\tilad}{{\tilde{\mrm a}}^\dagg}

\usepackage[backend=biber, natbib=true, style=numeric-comp, sorting=none, maxbibnames=99]{biblatex}
\def \scrip{{\cal I}^{+}}
\def \scrim{{\cal I}^{-}}
\def\e{\epsilon}

\def\a{\alpha}
\def\b{\beta}
\def\p{\partial}

\newcommand{\mrm}[1]{\mathrm{#1}}

\newcommand{\intinf}{\int_{-\infty}^{\infty}} 
\newcommand{\intsinf}{\int_{0}^{\infty}} 
\newcommand{\dagg}{\dagger}

\newcommand{\lie}{\mathcal{L}}
\newcommand{\cov}{\mathcal{D}}

\newcommand{\vac}{\ket{{\rm vac},\ \bar C_{ab} = 0, \ T_{cd} = 0 }}
\newcommand{\instate}{\ket{\rm in}}

\begin{document}    
\title{Generalized BMS algebra in higher even dimensions}
\author{Chandramouli Chowdhury$^{a}$, Anupam A. H.$^{a}$ and Arpan Kundu$^{b, c}$ }
\emailAdd{anupam.ah@icts.res.in,
chandramouli.chowdhury@icts.res.in, akundu@imsc.res.in}

\affiliation{$^a$International Centre for Theoretical Sciences, Tata Institute of Fundamental Research, Shivakote,
Bengaluru 560089, India.\\$^b$The Institute of Mathematical Sciences, Taramani, Chennai 600113, Tamil Nadu, India.
\\$^c$Homi Bhabha National Institute, Training School Complex, Anushaktinagar, Mumbai
400094, India.}

\abstract{We revisit the status of asymptotic symmetries in higher even dimensions and propose a definition of superrotation charge beyond linearized gravity. We prove that there is a well-defined spacetime action of the superrotation charge on the space of asymptotically flat geometries. Additionally, we demonstrate that the Ward identity associated with superrotation charges follows from the subleading soft graviton theorem, which is a universal constraint (in $d> 4$) along with the leading soft graviton theorem.}

\maketitle
\noindent
\flushbottom
\section{Introduction}
Starting with the seminal work by Strominger in 2013   \cite{Strominger:2013lka}, there has been a renewed interest in understanding the role of asymptotic symmetries in classical and quantum gravity.  This is primarily due to the fact that in $d = 4$ spacetime dimensions, the quantum gravity $\mathcal{S}$-matrix is constrained by a hierarchy of soft graviton theorems. The first two theorems in this hierarchy (namely, the leading and subleading soft graviton theorems) are universal, modulo the infrared loop effects. It was shown in a series of papers \cite{ Campiglia:2014yka,Campiglia:2015kxa, He:2014laa,Strominger:2014pwa,Campiglia:2015yka,Strominger:2013jfa,Kapec:2014opa} that these soft theorems are equivalent to the Ward identities associated with an algebra of infinite dimensional symmetries of the gravitational scattering. 

These symmetries form an infinite dimensional Lie algebra known as the extended or generalized BMS (eBMS/gBMS) algebra\footnote{In the literature, there are two possible candidate symmetry algebras, Ward identities of both of these symmetries are consistent with leading and subleading soft graviton theorems. These two algebras are referred to as the eBMS and gBMS algebras, respectively. For more details, we refer the reader to \cite{Campiglia:2014yka,Campiglia:2015yka,Barnich:2010eb,Barnich:2010ojg}}. For this paper, it is sufficient to review the gBMS algebra at null infinity as we shall only deal with massless excitations. This algebra is generated by {\it supertranslations} and smooth diffeomorphisms of the celestial plane, known as {\it superrotations}. The generators at future and past null infinities are identified via an antipodal identification map which ensures that one has a well-defined symmetry group of gravitational scattering where the incoming states are defined on the past boundary and outgoing states are defined on the future boundary of an asymptotically flat spacetime. 

In $d = 4$, one is finding increasing evidence that the gBMS algebra is a symmetry algebra of gravitational scattering and there are an infinity of charges which are conserved during the scattering process. In the classical theory, these charges define radiative observables (so called event-shapes) at null infinity, more commonly known as the memory. The supertranslation charge conservation is a statement about displacement memory effect \cite{Strominger:2014pwa} and the conservation of the superrotation charge produces the so called spin memory \cite{Pasterski:2015tva}.  Hence, one has a beautiful synthesis of symmetry, classical radiative observables and soft theorems, that constrains the gravitational $\mathcal{S}$-matrix in the infrared. 

In $d > 4$, the soft graviton theorems are more robust than in $d = 4$, as the 
$\mathcal{S}$-matrix is free of infrared divergences \cite{Campiglia:2014yka, Campiglia:2015kxa, Campiglia:2015yka}. However, it has been known since the works of Hollands and Wald  \cite{Hollands:2003xp, Hollands:2004ac} that classical gravity in higher dimensions has no displacement memory in the way it is defined in four dimensions. The same reasoning can be also used to argue that there can be no spin memory effect in higher dimensions.  This gives rise to an interesting puzzle in higher dimensions. The universality of soft theorems suggests that the $\mathcal{S}$-matrix is constrained by an infinite dimensional symmetry, but trivial event shapes at null infinity is seemingly at odd with infinitely many conserved charges. 
Recently, there has been much progress in resolving this tension.  It was first shown in  \cite{Kapec:2015vwa} that  in higher even dimensions in the  linearized regime, there exists a set of boundary conditions \cite{Hollands:2003xp,Hollands:2004ac, Kapec:2015vwa, Hollands:2003ie,Hollands:2013cva,Tanabe:2011es} such that, the corresponding solution space of (linearized) Einstein equations admit supertranslation symmetry. Consequently, the Ward identities corresponding to the supertranslation charges were shown to be equivalent to the leading soft graviton theorem \cite{Kapec:2015vwa}.  
Also, in \cite{Pate:2017fgt}, it was argued that the supertranslation charges in $d = 4 + 2k$ (where $k$ is any positive integer) define an observable, which is the higher dimensional analogue of the displacement memory. In \cite{Aggarwal:2018ilg} the supertranslation charges were put on a firmer footing by an ab-initio derivation of the same using the covariant phase space formalism.  In $d = 6$, it has been shown that  even in non-linear general relativity, supertranslation charges can be derived from the covariant phase space formalism and these are consistent with the known results in linearized gravity \cite{Chowdhury:2022nus}.

In this paper, we are interested in exploring the symmetry origin of the subleading soft graviton theorem in higher even dimensions. We show that the gBMS admits a natural extension to six dimensions in which the Poincar\'e group is enhanced to an infinite dimensional group composed of supertranslations and diffeomorphisms on the celestial plane $\mathbb R^{4}$ (superrotations). Superrotations in higher dimensions have been explored in several earlier works, and our analysis builds upon these results.  In \cite{Colferai:2020rte,Campoleoni:2020ejn,Campoleoni:2017qot}, the authors study superrotations in linearized gravity and the analysis in  \cite{Capone:2020mwy} focuses on understanding the set of boundary conditions that admit action of superrotations. 

In this paper, we derive the superrotation charge in six spacetime dimensions beyond the linearized regime.  In particular, we adopt a set of boundary conditions which allows the leading order angular metric $q_{ab}$ to fluctuate to any $u$-independent smooth metric \cite{Capone:2020mwy}.  As is well known, the choice of each such $q_{ab}$ is called the {\it frame}. The unit $\mathbb{R}^4$ metric is an example of the so called Bondi-frame, which will be defined rigorously in later sections. 

For frames that are infinitesimally away from the Bondi frame, we identify the correct radiative mode that transforms covariantly under superrotations. We  propose a definition of  the superrotation charge in the Bondi frame such that we get the aforementioned action on the radiative mode. We finally prove that the Ward identities corresponding to such superrotation charges follow from the subleading soft graviton theorem.

The paper is organized as follows. In section \ref{flatspace}, we review the boundary conditions that are adopted for asymptotically flat spacetimes in higher even dimensions. In section \ref{symgrav}, we show that the corresponding asymptotic symmetry algebra is the gBMS (which is a semidirect product of supertranslations and superrotations) and evaluate the spacetime action on the radiative phase space. We also identify the correct graviton mode in terms of the free data around the Bondi frame. In section \ref{SRCharges}, we propose the superrotation charges in the Bondi frame that generates the corresponding superrotation symmetry. In section \ref{wi}, we show that the Ward identities corresponding to the superrotation charges follow from the  subleading soft graviton theorem. We conclude and address about the future directions in section \ref{conclusion}. 

\section{Asymptotically flat spacetime in six dimensions}\label{flatspace}
In this section, we review asymptotically flat spacetimes in six dimensions. We shall analyse the corresponding asymptotic symmetries at null infinity in six dimensions in a modified version of the {\it Bondi gauge} \cite{Bondi:1962px,Sachs:1962wk}, which  shall be described below.

 Near future null infinity ${\cal I}^{+}$, the line element for the above class of spacetimes takes the following form
\be\label{bondi-1}
ds^2 = g_{\mu\nu} dx^\mu d x^{\nu} = M e^{2\b} du^2 - 2e^{2\beta} dudr + g_{ab} (dz^a - U^a du) (dz^b - U^b du)~,
\ee
where $u=t-r$ is the retarded time, $r$ is the radial distance and $z^a$ denotes the coordinate on the celestial plane\footnote{This is equivalent to the decompactified celestial sphere. The coordinate transformation from the Bondi coordinates to these can be found in \cite{Kapec:2017gsg}. All formulas written in this paper till section 3 trivially generalize to the celestial sphere.} $\mathbb R^4$. Note that the index of $U_a$ are lowered and raised using the metric $g_{ab}$. 

The components $M, \beta, U^a$ and $g_{ab}$ in \eqref{bondi-1} are functions of the $(u, r, z^a)$ coordinates and they have the following radial expansions near $\scrip$ \cite{Kapec:2015vwa},
\begin{eqn}\label{bondi-2}
	M &= \sum_{n = 0}^{\infty} \frac{M^{(n)}(u, z)}{r^n}, \qquad 
	\beta = \sum_{n = 2}^\infty \frac{\beta^{(n)}(u, z)}{r^n}, \qquad 
	U_a = \sum_{n = 0}^\infty \frac{U_a^{(n)}(u, z)}{r^n}, \\
	g_{ab} &= r^2 q_{ab}(z) + \sum_{n = -1}^{\infty} \frac{g_{ab}^{(n)}(u, z)}{r^n} \equiv r^2 q_{ab}(z) + r C_{ab}(u, z) + D_{ab}(u, z) + \frac{E_{ab}(u, z)}{r} + \frac{F_{ab}(u, z)}{r^2} + \cdots
\end{eqn}

We consider the space of asymptotically flat geometries where the metric on the celestial plane, $q_{ab}$, is chosen to be independent of $u$. The interested readers can refer to \cite{Capone:2021ouo} for generalizations to $u$-dependent $q_{ab}$. The indices of the components $U_a^{(n)},\ g_{ab}^{(n)}$ are lowered and raised using $q_{ab}$. Along with the expansion \eqref{bondi-1}, there is an additional gauge fixing condition, often referred to as the {\it Bondi determinant condition} which is given as
\be\label{bondi-det}
\det\Big(\frac{g_{ab}}{r^2} \Big) = \det(q_{ab}) = \det(\delta_{ab})~,
\ee
where $\delta_{ab}$ is the metric on $\mathbb R^4$. We would like to point out a key difference between the leading order angular metric chosen in our paper (denoted by $q_{ab}$) with those chosen in earlier literature \cite{Kapec:2015vwa,  Aggarwal:2018ilg, Chowdhury:2022nus}. In previous works \cite{Kapec:2015vwa,  Aggarwal:2018ilg, Chowdhury:2022nus}, this metric was either fixed to be the unit sphere metric $\mathbb S^4$ ($\gamma_{ab}$) or the metric on the plane $\mathbb R^4$ ($\delta_{ab}$) and further analysis of asymptotic symmetries were pursued with this choice. This led to the proposal for the asymptotic symmetry group as the six dimensional BMS group, which is the semi-direct product of supertranslations and the Lorentz group $SO(5,1)$.

As will be shown, just as in four dimensions, relaxing the metric on the celestial plane to an arbitrary smooth metric (with the determinant condition \eqref{bondi-det}) leads to an extension of the BMS algebra in six dimensions, that we refer to as the generalized BMS algebra. In four dimensions, the choice $q_{ab}=\delta_{ab}$  is referred to as the {\it Bondi frame}. However in six and higher dimensions, the Bondi frame can be understood as the choice for $q_{ab}$ which satisfies the four dimensional Einstein equation  (with or without a cosmological constant)\footnote{For example, the cosmological constant is needed when $q_{ab} = \gamma_{ab}$ (the metric of the unit sphere) but is not needed when $q_{ab} = \delta_{ab}$ (the metric of the unit plane).}. A metric $q_{ab}$ in the non-Bondi frame does not satisfy the four dimensional Einstein equation. However, in four spacetime dimensions, a similar definition does not apply to the corresponding two dimensional angular metric $q_{ab}$ on the celestial plane/sphere. For our purposes, the Bondi frame in six dimensions shall be always referred to as $q_{ab} = \delta_{ab}$.

Using the determinant condition \eqref{bondi-det}, it can be shown that the traces of $g_{ab}^{(n)}$ are fixed in terms of $g_{ab}^{(n-1)}$. For example, 
\begin{eqn}\label{trace-conditions}
	C_a^a &= 0, \\
	D^a_a &= \frac{1}{2} C_{ab} C^{ab}, \\
	E^a_a &= C^{ab} D_{ab} - \frac{1}{3} C^{am} C_{mn} C^n_a, \\
	F^a_a &= C^{ab} E_{ab} + \frac{1}{2} D^{ab} D_{ab} - C^{am} C_{mn}  D^n_a  + \frac{1}{4} C^{am} C_{mn} C^{nb} C_{ba}~.
\end{eqn}
Having expressed the general form of an asymptotically flat spacetime, we can now solve the Einstein equations for the above family of metrics. This also requires us to impose the following fall off conditions on the Ricci tensor, which are motivated by demanding the finiteness of energy flux and other physical observables \cite{Kapec:2015vwa},
\begin{eqn}\label{ricci-falloffs}
	R_{uu} &= O(r^{-4}), \qquad R_{ur} = O(r^{-5}), \qquad R_{ua} = O(r^{-4}), \\
	R_{rr} &= O(r^{-6}), \qquad R_{ra} = O(r^{-5}), \qquad R_{ab} = O(r^{-4})~.
\end{eqn}
Using the equations above, we find that all metric components in \eqref{bondi-2} can be expressed in terms of $q_{ab}, C_{ab}$ and $D_{ab}$. For example, it can be shown that,
\begin{alignat}{2}\label{eom}
	M^{(0)}&=-\frac{\bar{\mathcal{R}}}{12}, \qquad  &&U_{a}^{(0)}=-\frac{1}{6}\cov_{b}C_{a}^{b}~,\\
	\beta^{(2)}&=-\frac{1}{64}C^{ab}C_{ab}, \qquad  &&\beta^{(3)}= \frac{1}{48}\Big(C^{ab}C_{bm}C^{m}_{a} - 2C^{ab}D_{ab}\Big)~,
\end{alignat}	
where $\bar{\mathcal{R}}$ is the Ricci scalar for the leading order angular metric $q_{ab}$ and $\cov_a$ denotes the covariant derivative w.r.t $q_{ab}$. The Einstein equations also imposes the following condition on $C_{ab}(u, z)$,
\begin{align}\label{cab}
	\partial_{u} C_{ab}(u, z)=-\bar{\mathcal{R}}_{ab}^{\tf} \equiv -\bar{\mathcal{R}}_{ab} + \frac{1}{4}q_{ab}\bar{\mathcal{R}}~,
\end{align}
where $\bar{\mathcal{R}}_{ab}$ is the Ricci tensor w.r.t $q_{ab}$. This implies that the general solution for $C_{ab}$ can be written as
\begin{align}\label{cab-full}
	C_{ab}(u,z)=\bar{C}_{ab} (z)+u T_{ab} (z)~,
\end{align}
where 
\begin{align}\label{tab}
	T_{ab} = -\bar{\mathcal{R}}_{ab}^{\tf}~.
\end{align}
We conclude this section with a few remarks.
\begin{itemize}
\item $D_{ab}(u,z)$ is the unconstrained dynamical  data in six dimensions, i.e, it is not determined by the equations of motion. 

\item In the Bondi frame, $T_{ab} = 0$.

\item In $d = 4$, the physical News tensor (which encodes gravitational radiation) is determined by subtracting the {\it Schouten Tensor} at $\scrip$ from the {\it Geroch Tensor}, $\Phi_{ab}$. $\Phi_{ab}$ is determined by the requirement that in any frame at $\scrip$, the News is gauge invariant under the unphysical Weyl rescaling at $\scrip$. There is a well-defined relationship between $\Phi_{ab}$ and the tensor $T_{ab}$ in $d = 4$ \cite{Campiglia:2020qvc}. However, in six dimensions, the analogous relationship between $T_{ab}$ and $\Phi_{ab}$ is not clear and is beyond the scope of this paper. For an earlier discussion of this issue, we refer the reader to \cite{Capone:2021ouo}.  

 \end{itemize}

\section{Generalized BMS in six dimensions}\label{symgrav}
In this section, we review the  asymptotic symmetries at null infinity in six dimensions.
The asymptotic symmetry group associated to the class of metrics described in the previous section is the group of diffeomorphisms that preserve the form of the metric at $\scrip$ \eqref{bondi-1} and also satisfies the determinant condition \eqref{bondi-det}. Generators of such diffeomorphisms are vector fields which are divergence free  at $\scrip$ \eqref{cond3}.

 \subsection{Generators of supertranslations and superrotation}
Consider a smooth vector field $\xi$ of the following form
\begin{align}\label{killing1}
	\xi = \xi^{u}(u,r,z)\partial_{u} +
	\xi^{r}(u,r,z)\partial_{r} + \xi^{a}(u,r,z)\partial_{a}~.
\end{align}
Gauge fixing conditions \eqref{bondi-1} together with \eqref{bondi-det} imply that the vector fields have to satisfy 
\begin{align}\label{cond1}
	\mathcal{L}_{\xi}g_{rr}=0,\qquad \mathcal{L}_{\xi}g_{ra}=0,\qquad g^{ab}\mathcal{L}_{\xi}\det g_{ab}&=0~.
\end{align}
The above conditions fixes the form of the vector fields to be
\begin{subequations}
\begin{align}\label{gBMSvv1}
	\xi^u(u, r, z) &= W(u,z)~, \\\label{gBMSvv2}
	\xi^a(u, r, z) &= V^a(z) - \cov_b W(u,z) \int_r^{\infty}  e^{2\beta(r')} g^{ab}(r') dr'~, \\\label{gBMSvv3}
	\xi^r(u, r, z) &= - \frac{r}{4} \big[ \cov_a \xi^a(u, r, z) - U^a \cov_a W(u,z) \big]~,
\end{align}
\end{subequations}
where  $W(u,z)$ is an arbitrary function on the celestial plane, and $V^{a}(z)$ is a smooth vector field on the same. The vector field $\xi$ can be determined by the fall off conditions given in \eqref{bondi-2} along with the divergence free condition, which are stated as,
\begin{subequations}
\begin{align}\label{cond2}
	\mathcal{L}_{\xi}g_{uu}=\mathcal{O}(1),\qquad &\mathcal{L}_{\xi}g_{ur}=\mathcal{O}(r^{-1}),\qquad \mathcal{L}_{\xi}g_{ab}=\mathcal{O}(r^2)~,\\
	&\lim\limits_{r\rightarrow\infty}\nabla_{\mu}\xi^{\mu}=0~.\label{cond3}
\end{align}
\end{subequations}
Here $\nabla$ denotes the covariant derivative w.r.t to the metric $g_{\mu\nu}$ given in \eqref{bondi-1}. Using the conditions above, $W(u,z)$ is fixed as
\begin{align}
	W(u,z)=f(z)+u\alpha(z)~,
\end{align}
where $f(z)$ is an arbitrary smooth function on the celestial plane, and $\a = \frac{1}{4}\cov_a V^a$. Thus, one can see that the vector field $\xi$ is parameterized by $f(z)$ and $V^a(z)$. The vector fields characterised by $f(z)$ (by setting $V^a(z) =0$) are called the supertranslation vector fields while the vector fields characterised by $V^a(z)$ (by setting $f(z)=0$) are called the superrotation vector fields. Therefore, the supertranslation vector field $\xi_{f}$ can be written as \cite{Chowdhury:2022nus} 
\begin{subequations}
\begin{align}
\label{st1}
	\xi_f^u(u, r, z) &= f(z)~, \\\label{st2}
	\xi_f^a(u, r, z) &=  - \cov_b f(z) \int_r^{\infty}  e^{2\beta(u,r',z)} g^{ab}(u,r',z) dr'~, \\\label{st3}
	\xi_f^r(u, r, z) &= - \frac{r}{4} \big[ \cov_a \xi_f^a(u,r,z) - U^a(u,r,z) \cov_a f (z)\big]~.
\end{align}
\end{subequations}
The superrotation vector field $\xi_V$ can be also written as
\begin{subequations}
\begin{align}
\label{sr1}
	\xi^u_V(u, r, z) &=  u \a(z)~, \\\label{sr2}
	\xi^a_V(u, r, z) &= V^a(z) -  u \cov_b\a(z) \int_r^{\infty}  e^{2\beta(u,r',z)} g^{ab}(u,r',z) dr'~, \\\label{sr3}
	\xi^r_V(u, r, z) &= - \frac{r}{4} \big[ \cov_a \xi_V^a(u,r,z) - u U^a(u,r,z) \cov_a \a(z)  \big]~.
\end{align}
\end{subequations}
Hence, the gBMS algebra is defined as the asymptotic symmetry algebra generated by the supertranslation vector field ($\xi_f$) and the superrotation vector field ($\xi_V$). For the purpose of this paper, we mainly focus our attention to superrotations. The details of the symmetry algebra and corresponding charge algebra is left for a future work.

 \subsection{Spacetime action on radiative phase space}
Using \eqref{killing1}, we can derive the action of supertranslations and superrotations on the variables parameterizing the phase space. We note that the background metric $q_{ab}$ remains invariant under supertranslations but transforms under the action of superrotations\footnote{We use the notation $\hat\delta_f$ and $\hat\delta_V$ to denote the variations computed by setting $V^a = 0$ and $f = 0$ in $\hat \delta_\xi$ respectively in any general frame.},
\begin{eqn}\label{qabvar}
\hat\delta_f q_{ab} &= 0~, \\
\hat\delta_V q_{ab} &= - 2 \a q_{ab} + \lie_V q_{ab} = - 2 \a q_{ab} + 2 q_{c(a} \cov_{b)} V^c~,
\end{eqn}
where we use the symmetrization convention $X_{(a} Y_{b)} = \frac{1}{2}(X_a Y_b + X_b Y_a)$. Using \eqref{qabvar}, it is easy to see that under the action of superrotation, a Bondi frame ($T_{ab} = 0$) generically transforms to a non-Bondi frame ($T_{ab} \neq 0$). Upon using a stronger fall off condition, $\hat \delta_V g_{ab} = O(r)$, we get a constraint on $V^a$, which takes the form of a conformal Killing vector (CKV) equation, 
\be
	\mathcal{D}_{a}V_b + \mathcal{D}_b V_a -  \frac{q_{ab}}{2} \mathcal{D}_c V^c =0~.
\ee
The solutions to the CKV equation above are the generators of Lorentz transformations which are finite dimensional. Hence by imposing  less restrictive fall offs, we allow an infinite dimensional extension of the Lorentz group in six dimensions\footnote{This is similar to the four dimensional case. In four spacetime dimensions there are two extensions possible, which is the extended BMS \cite{Barnich:2010eb} and generalized BMS\cite{Campiglia:2015kxa,Campiglia:2015yka}. Extended BMS group is generated by $V^{a}$'s which are local CKV's in two dimensions. In six dimensions, this extension is not possible as the solution to the CKV is finite dimensional.}.

We now discuss the action of gBMS transformations on the radiative phase space, i.e, $C_{ab}(u, z) = \bar C_{ab}(z) + u T_{ab}(z)$ and $D_{ab}(u, z)$. These can be derived by studying the variation $\hat \delta_\xi g_{ab}$ and expanding it in powers of $r$. The action of supertranslations gives,
\begin{eqn}\label{STvar1}
\hat \delta_f \bar C_{ab} &= \frac{1}{2} \cov^2 f q_{ab} - 2 \cov_a \cov_b f + f T_{ab}, \\
\hat \delta_f T_{ab} &= 0, \\
\hat \delta_f D_{ab} &= f \p_u D_{ab} + \frac{1}{4} \cov^2 f C_{ab} - U_{(a}^{(0)} \cov_{b)} f + \frac{1}{2} q_{ab} U^{(0)c} \cov_c f - \frac{1}{4} q_{ab} \cov_c(C^{cd} \cov_d f)  \\
&\quad - \frac{1}{2} C_{c(a)} \cov_{b)} \cov^c f - \cov^c f \cov_c C_{ab} + \frac{1}{2} \cov^c f \cov_{(a} C_{b) c} ~.
\end{eqn}
These equations generalize the action of supertranslations on the phase space variables in a non-Bondi frame. Upon setting $T_{ab} = 0$ (Bondi frame)  we recover the results in \cite{Chowdhury:2022nus}. 

Note that $T_{ab}$ is invariant  under supertranslations as $\hat\delta_f q_{ab}$ = 0 (see \eqref{tab} and \eqref{qabvar}). The action of superrotations on the radiative phase space can be derived in a similar manner,
\begin{eqn}\label{SRvar1}
	\hat \delta_V \bar C_{ab} &= \lie_V \bar C_{ab} - \a \bar C_{ab}~, \\
	\hat \delta_V T_{ab} &= \lie_V T_{ab}  - 2  \big( \cov_a \cov_b \a \big)^{\tf}~, \\
 \hat \delta_V D_{ab} &= u \a \p_u D_{ab} + \lie_V  D_{ab} \\
&\quad+ u \Bigg\{ \frac{1}{4} \cov^2 \a  C_{ab} -  U_{(a}^{(0)} \cov_{b)} \a + \frac{1}{2} q_{c(a} \cov_{b)} \big( C^{cd} \cov_d \a \big) - C_{c(a} \cov_{b)} \cov^c \a \\
&\qquad - \cov^c \a \cov_c C_{ab}  + \frac{1}{2}q_{ab} U^{(0)c} \cov_c \a - \frac{1}{4}q_{ab} \cov_c (C^{cd} \cov_d \a)  \Bigg\} ~.
\end{eqn}
 The second equation above can be independently derived by evaluating the variation $ \hat \delta_V \bar{\mathcal R}_{ab}$. Note that the variation of $D_{ab}$ in the equations above are expressed in terms of $C_{ab} = \bar C_{ab} + u T_{ab}$ for ease of notation. 
 In the rest of this section and the following (section \ref{SRCharges}), we analyze the action of gBMS symmetries on the sector of the radiative phase space where $T_{ab} = 0$. We note that even though this sector is preserved under the action of infinitesimal supertranslations, under infinitesimal superrotations any configuration in the $T_{ab} = 0$ sector is generically mapped to a configuration where $T_{ab} \neq 0$. A general analysis of gBMS symmetries on the full radiative phase space at $\scrip$ is beyond the scope of this paper and is left for future work.

The variations \eqref{SRvar1} take a simpler form in the Bondi frame, where we have to set $T_{ab} = 0$\footnote{We would like to draw attention to the notational differences between $\delta$ and $\hat \delta$. The latter refers to a variation in any general frame, whereas the former is a variation specifically evaluated in the Bondi frame ($T_{ab} = 0$).},
\begin{eqn}\label{SRvar3}
	\delta_V \bar C_{ab} &= \lie_V \bar C_{ab}  - \a \bar C_{ab}~, \\
	\delta_V T_{ab} &=- 2  \big( \partial_a \partial_b \a \big)^{\tf}~, \\
	\delta_V D_{ab} &= u \a \p_u D_{ab} + \lie_V  D_{ab} \\
	&+ u \Bigg\{ \frac{1}{4} \partial^2 \a \bar C_{ab} -  U_{(a}^{(0)} \partial_{b)} \a + \frac{1}{2} q_{c(a} \partial_{b)} \big( \bar C^{cd} \partial_d \a \big) - \bar C_{c(a} \partial_{b)} \partial^c \a \\
	&\qquad - \partial^c \a \partial_c \bar C_{ab}  + \frac{1}{2}q_{ab} U^{(0)c} \partial_c \a - \frac{1}{4}q_{ab} \partial^c (\bar C_{cd} \partial^d \a)  \Bigg\}~.
\end{eqn}
From \eqref{SRvar3}, it is clear that the dynamical data $D_{ab}$ in the Bondi frame grows as $O(|u|^1)$ as we take $|u| \to \infty$. Therefore, $D_{ab}$ by itself does not represent the graviton mode (in a frame where $\bar{C}_{ab} \neq 0$) since this is in contradiction with the expected fall off from the saddle point analysis and computation of the symplectic form \cite{Chowdhury:2022nus} where one gets
\be\label{physicalfalloff}
\lim_{|u| \to \infty} {\rm Graviton} \sim  O\left( \frac{1}{|u|^{2+0^+}}\right)~.
\ee
 This was already noticed in \cite{Chowdhury:2022nus} for the case of supertranslations in the Bondi frame, where the authors identified the graviton mode for $\bar C_{ab} \neq 0$ as a redefinition of $D_{ab}$, given as,
 \be\label{Dtil1}
 D_{ab} \to \tilde D_{ab}^{ST} = D_{ab} - \frac{1}{4} \bar C_{a}^m \bar C_{bm} - \frac{1}{16} \delta_{ab}\bar  C_{mn}\bar C^{mn}~.
\ee 
Using the following form for $\bar C_{ab}$ in the Bondi frame,
\begin{eqn}
	\bar C_{ab} = - 2 \Big( \partial_a \partial_b \psi \Big)^{\tf}~,
\end{eqn}
which follows from the  vanishing of the Weyl tensor $C_{urab}(u = \pm\infty, z)$ at $O(r^{-1})$ \cite{Kapec:2015vwa, Aggarwal:2018ilg}, the action of supertranslation on $\tilde D_{ab}^{ST}$ is given as,
\be
\delta_f\tilde D_{ab}^{ST} = f \p_u \tilde D_{ab}^{ST}~.
\ee

Surprisingly, the same redefinition but with $\delta_{ab} \to q_{ab}$ and $\bar C_{ab} \to  \bar C_{ab} + u T_{ab}$ ensures that for linear deviations from the Bondi frame, we get proper fall offs for superrotated fields, i.e,
\begin{eqn}\label{Dtilvariations}
 \delta_f \tilde D_{ab}  &= f \p_u \tilde D_{ab}~, \\
\delta_V \tilde D_{ab} &= \lie_V \tilde D_{ab} + u \a \p_u \tilde D_{ab}~,
\end{eqn}
where 
\begin{eqn}\label{SRcorrectRedefn}
\tilde D_{ab} &= D_{ab} - \frac{1}{4} q^{mn} \bar C_{am} \bar C_{bn} - \frac{1}{16} q_{ab}\bar C_{mn} \bar C^{mn} \\
&\quad - u \Big[ \frac{1}{4} q^{mn} (\bar C_{am} T_{bn} + T_{am} \bar C_{bn}) + \frac{1}{8} q_{ab} T_{mn} \bar C^{mn} \Big] + O(T^2)~,
\end{eqn}
with the fall off condition
\be\label{fall off}
\lim_{u \to \pm \infty}\tilde D_{ab} = O\Big( \frac{1}{u^{2+0^+}} \Big)~.
\ee
By using the redefined field \eqref{SRcorrectRedefn} the News tensor $\p_u \tilde D_{ab} = \p_u D_{ab}$ is unchanged in the Bondi frame as $T_{ab} = 0$. One might be worried  that even though we are finally working in the Bondi frame, it is necessary to include $T_{ab}$ in the definition above \eqref{SRcorrectRedefn}. This can be explained as follows. From \eqref{SRvar1} it is clear that $T_{ab}$ transforms non-homogeneously, i.e, even if one starts in the Bondi frame ($T_{ab}=0$), under superrotations $T_{ab}$ transforms to $- 2  \big( \cov_a \cov_b \a \big)^{\tf}$. 
The terms which are linear in the redefinition above will ensure that the non-homogeneous terms generated from the variation of $D_{ab}$ get appropriately cancelled with the non-homogeneous terms generated by the variation of $T_{ab}$, which is essential in order to respect the fall off condition. In order to derive the generic form for $\tilde D_{ab}$ with appropriate fall off conditions in a general non-Bondi frame, we need to take care of the $O(T^2)$ terms in \eqref{SRcorrectRedefn} which is beyond the scope of this paper.

Equations \eqref{Dtilvariations}, \eqref{SRcorrectRedefn} are among the central results of the paper as they display the correct phase space variables to use in the Bondi frame in the presence of both supertranslations and superrotations.
 \subsection{Generalized BMS at $\scrim \cup \scrip$}
The gBMS symmetry algebra at $\mathcal{I}^+$ (denoted by $\mathcal{G}^+$) is defined as the symmetry algebra generated by supertranslations and superrotations on the radiative phase space at $\mathcal{I}^+$. Similarly, one can independently define the asymptotic symmetry algebra at $\mathcal{I}^-$ (denoted by $\mathcal{G}^-$). In order to define a gravitational scattering problem that takes the incoming scattering data at $\mathcal{I}^-$ to outgoing scattering data at $\mathcal{I}^+$, one must define a common asymptotic symmetry algebra at $\scrim \cup \scrip$. 
Motivated from \cite{Campiglia:2014yka,Campiglia:2015yka}, where the analysis was performed in four dimensions, it is natural to propose that in six dimensions, the diagonal subalgebra of gBMS is the symmetry algebra of the quantum gravity $\mathcal{S}$-matrix. The diagonal subalgebra is identified using the  antipodal matching conditions on the null generators of $\mathcal{G}^+$ and $\mathcal{G}^-$ which are given as
 \begin{eqn}
 f_{+}(z) &= f_{-}(-z)~,\\
 V_{+}^a(z) &= V_{-}^a(-z)~.
 \end{eqn}
 Here, $ \big( f_{+} , V_{+}^a\big) , \big(f_{-},V_{-}^a\big)$  denote the parameterizations used for supertranslations and superrotations  at $\mathcal{I}^+$ and $\mathcal{I}^-$ respectively. 

\section{Superrotation charge in the Bondi frame}\label{SRCharges}
In this section, we derive the superrotation charge in the Bondi frame  that generates the spacetime action on the radiative phase space.  As in four dimensions, the superrotation charge consists of two independent terms which we refer to as the soft and the hard charge respectively\footnote{The nomenclature is motivated by analysis of these charges in four dimensions where the soft superrotation charge is the so-called spin memory \cite{Pasterski:2015tva}.}. We remind the readers that the metric at the leading order in large$-r$ in the Bondi frame is
\be
ds^2 =  -2du dr + r^2 \delta_{ab} dz^a dz^b~.
\ee
With this choice, we shall proceed onto computing the charges corresponding to the asymptotic symmetries\footnote{Even though we are working with the metric on the decompactified sphere, none of the physical outcomes will change by considering the metric on the unit sphere.} discussed in section \ref{symgrav}. Computing the charges using the Noether procedure requires a thorough understanding of the symplectic structure in a non-Bondi frame, which is outside the purview of this paper. However, the same can be used to compute the hard charge even in this case, but obtaining the total superrotation charge is difficult. Therefore, we shall adopt an alternative route to obtain the charge where we exploit the connection between the soft theorem and the Ward identities (corresponding to the charges). 

The charges we obtain by this method can be easily generalized to gravity coupled to any spin field. Specifically, we shall consider the special case of the gravity coupled to scalars and explicitly demonstrate the equivalence of the Ward identity and the subleading soft theorem in this example. Our work is based on a similar approach by the authors in \cite{Colferai:2020rte}. We notice certain subtleties associated with their analysis which are delineated and improved upon in the upcoming sections.  

\subsection{Soft charge}
Motivated by the structure of the soft charge in 4-dimensions, we write down the general tensor structure that is covariant and also generates the correct transformation for the radiative data $\bar C_{ab}$ (the necessary Poisson brackets can be read from the symplectic structure given in \cite{Chowdhury:2022nus}). We propose that with these conditions, the soft charge is given as 
\begin{eqn}\label{softcharge}
Q^S_V =  \frac{64\pi^2}{128 \pi G_N}\intsp u V^b(x) \mathbb D^a \tilde D_{ab}  + \frac{1}{96 \pi G_N} \int_{\scrip} (\lie_V \bar C_{ab} - \a \bar C_{ab}) \partial^a \partial^m \tilde D^b_m~,
\end{eqn}
where we use the notation $\intsp \equiv \intinf du \int d^4 z$ to denote integrals over $\scrip$ and the derivative operator $\mathbb D^a \tilde D_{ab}$ is given as \cite{Kapec:2017gsg}
\be
\mathbb D^a \tilde D_{ab} =  \frac{1}{64\pi^2} \Big[ \p^4 \p^a \tilde D_{ab} - \frac{4}{3} \p_b \p^2 \p^{ef} \tilde D_{ef}   \Big].
\ee
The first term in the soft charge \eqref{softcharge} is derived in \cite{Kapec:2017gsg} by relating it to the CFT$_4$ stress tensor on the boundary and also in \cite{Banerjee:2019tam} with arguments based on the conformal properties of such operators. In the upcoming section, we demonstrate how this is consistent with the subleading  soft theorem for gravitons coupled to scalars. Although this has been derived in a specific frame, $\bar C_{ab} = 0$ (where $\tilde D_{ab} = D_{ab}$), as seen from the sections before, the correct variable to use in a $\bar C_{ab} \neq 0$ frame is $\tilde D_{ab}$ and hence the first term is a generalization of the result in \cite{Kapec:2017gsg} to a general supertranslated Bondi frame. The second term in \eqref{softcharge} is new and follows by demanding that the charge generates the correct spacetime transformations for $\bar C_{ab}$ in the Bondi frame. This requires us to use the Poisson bracket derived in \cite{Kapec:2015vwa, Chowdhury:2022nus}, 
\be
\left\{\intinf du\ \p^2 \p^{ab} \tilde D_{ab}(u, z),\ \psi(z')\right\} = 96\pi G_N \delta(z, z')~.
\ee
 Note that a derivation of $Q_V^S$ from a purely asymptotic symmetry perspective requires us to study the symplectic structure carefully. This has been carried out in four dimensions \cite{Campiglia:2015yka}, and extending to higher dimensions is currently a work in progress. 

We would like to point out that the expression for the soft charge \eqref{softcharge}, differs from the one given in \cite{Colferai:2020rte}. In order to compare the two expressions we first set $\bar C_{ab}$ in \eqref{softcharge} as this is the case studied in \cite{Colferai:2020rte}. This will leave us with only the first term in \eqref{softcharge} which upon expanding gives,
 \bes
 \frac{1}{128\pi G_N}\intsp u V^b(z) \Big[ \p^4 \p^a D_{ab} - \frac{4}{3} \p_b \p^2 \p^{ef} D_{ef}   \Big].
 \ees
  After performing an integration by parts, we see that the second term in the expression above is proportional to $\a$, and matches with the soft charge proposed in \cite{Colferai:2020rte} upto a proportionality factor. However, as will be shown in the next section, if the first term is not included in the soft charge, there is a inconsistency from the perspective of subleading soft graviton theorem\footnote{In \cite{Colferai:2020rte}, the commutator of the charge with the radiative data was studied upto a proportionality factor and hence the extra term might have been missed.}.

\subsection{Hard charge}
Having discussed the soft charge in the previous section, we now derive the gravitational superrotation hard charge  by two methods. The first method employs the gravitational stress energy tensor derived in \cite{Laddha:2019yaj} and will be explained in this section. The same expression for the hard charge can also be derived using the symplectic form for hard sector of the radiative phase space as shown in appendix \ref{app:sympform-hardcharge}.

 Using the gravitational stress energy tensor, the supertranslation \cite{Chowdhury:2022nus} and superrotation hard charge are proposed as,
\begin{subequations}\label{hardcharge}
	\be\label{SThard1}
\big. Q_f^{H} \big|_{\bar C = 0} = \frac{1}{8\pi G_N} \intsp f \stress_{uu}~, 
\ee
\be\label{SRhard1}
\big. Q_V^{H} \big|_{\bar C = 0}	= \frac{1}{8\pi G_N} \intsp  u \a \stress_{uu} + V^a \stress_{ua} ~,
\ee
\end{subequations}
where the stress tensors are given as \cite{Laddha:2019yaj}, 
\begin{subequations}
\be\label{STstress}
	\stress_{uu} = \frac{1}{4} N_{ab} N^{ab}, 
	\ee
\be\label{SRstress}
	\stress_{ua} = \frac{1}{4} \Big[  N_{bc} \partial_a D^{bc} - 2 N^{bc} \partial_c D_{ab} + 2 N_{ca} \partial_b D^{bc}   \Big].
\ee
\end{subequations}
The notation $\big. Q \big|_{\bar C =0}$ emphasizes the fact that the background used for the computations above is the usual flat metric without turning on supertranslations.

Upon simplifying the expressions above, the hard charges can be written as
\begin{subequations} 
\be\label{SThard2}
\big. Q_f^H  \big|_{\bar C = 0}= \frac{1}{32\pi G_N} \intsp \ f(z) N_{ab} N^{ab},
\ee
\be\label{SRhard2}
\big. Q_V^H \big|_{\bar C = 0} = \frac{1}{32\pi G_N} \intsp \ N^{ab} \Big( \lie_V D_{ab} + u \a N_{ab} \Big)~.
\ee
\end{subequations}
As demonstrated previously, the correct phase space variable governing the radiative data in the Bondi frame (with $\bar C_{ab} \neq 0$) is $\tilde D_{ab}$ and using the fall off condition \eqref{fall off}, the form of the hard charge is unchanged,
\begin{subequations} 
	\be\label{SThard3}
	Q_f^H = \frac{1}{32\pi G_N} \intsp\ f(z) N_{ab} N^{ab},
	\ee
	\be\label{SRhard3}
	Q_V^H  = \frac{1}{32\pi G_N}  \intsp\ N^{ab} \Big( \lie_V \tilde D_{ab} + u \a N_{ab} \Big).
	\ee
\end{subequations}
This charge can also be derived by analyzing the hard sector of the symplectic structure \cite{Campiglia:2021bap}. The symplectic structure has been derived in \cite{Chowdhury:2022nus} by working with $\delta q_{ab}= 0$, which is sufficient for the purpose of deriving the hard charge for both supertranslation and superrotation. The supertranslation hard charge \eqref{SThard3} has been derived in \cite{Chowdhury:2022nus} and in appendix \ref{app:sympform-hardcharge} we describe how the superrotation hard charge can be derived from the symplectic structure which matches with \eqref{SRhard3}.

\subsection{Total superrotation charge}
The total superrotation charge in the Bondi frame is given as the sum of \eqref{softcharge} and \eqref{SRhard3},
\begin{eqn}\label{totalSRcharge}
Q_V &= 	 \frac{1}{32\pi G_N} \int_{\scrip} N^{ab} \Big( \lie_V \tilde D_{ab} + u \a N_{ab} \Big) \\
&\quad+ \frac{\pi}{2 G_N}\intsp u V^b(x) \mathbb D^a \tilde D_{ab}  + \frac{1}{96 \pi G_N} \int_{\scrip} (\lie_V \bar C_{ab} - \a \bar C_{ab}) \partial^a \partial^m \tilde D^b_m~.
\end{eqn}
Using the symplectic form \eqref{hardsym}, it can be easily shown that the Poisson bracket between the hard charge and the radiative data $\tilde D_{ab}(u, z)$  reproduces the superrotation spacetime action, i.e, $\delta_V \tilde D_{ab}(u, z)$.

\section{Ward identity and subleading  soft graviton theorem}\label{wi}
In six dimensions, both the leading and subleading soft graviton theorems are exact constraints on $\mathcal{S}$-matrix of quantum gravity \cite{Sen:2017xjn}. In four dimensions, any statement on the $\mathcal{S}$-matrix has to be understood with care as the Dyson $\mathcal{S}$-matrix is infrared divergent. However in higher dimensions, soft theorems are precise factorisation statements about the $\mathcal{S}$-matrix which is infrared finite. Thus a relationship of the asymptotic symmetries with the soft theorems is likely to be a rather robust statement about asymptotic symmetries of the $\mathcal{S}$-matrix even when loop effects are taken into account. In this section, we will argue that the subleading quantum soft graviton theorem with external states being massless scalars in six dimensions imply the Ward identity for $\textrm{Diff}(\mathbb R^{4})$ asymptotic symmetries. The similar analysis with finite energy external gravitons (or any other non-zero spin) requires a careful understanding of quantization of the gravitons in a non-Bondi frame  we leave for future work. 

We begin with the quantization of the soft charge in the Bondi frame. Using the saddle point approximation, the mode expansion of the graviton in the Bondi frame is given as\footnote{There is an analogous mode expansion for a field with spin$-s$ in six dimensions,
\be
\tilde X_{m_1 \cdots m_s}(u, \hat x) \propto \intsinf d\omega  \omega \Big[ \tila_{m_1 \cdots m_s}(\hat x) e^{- i \omega u}+ \tilad_{m_1 \cdots m_s}(\hat x)  e^{i \omega u} \Big].
\ee} (these formulas are derived in great detail in \cite{Kapec:2017gsg, He:2019jjk}, where in order to match with their conventions we need to replace $u \to \frac{u}{2}$ in our formulas)
\be\label{Dabquant}
 \tilde D_{ab}(u, \hat z) = - \frac{\sqrt{8\pi G_N}}{(2\pi)^3} \intsinf d\omega  \omega \Big[ \tila_{ab}(\omega, \hat z) e^{- i \omega u}+ \tilad_{ab}(\omega, \hat z)  e^{i \omega u} \Big]~,
\ee
where $\tila_{ab}(\omega, \hat z)$ and $\tilad_{ab}(\omega, \hat z)$ are the  annihilation and creation operators for the graviton in the vacuum labelled by $\bar C_{ab}$, respectively (see section \ref{sec:vacuum}). They satisfy the following commutation relation 
\be
 [\tila_{ab}(\omega_1, z_1), \tilad_{cd}(\omega_2, z_2)] = \frac{2(2\pi)^5}{\omega_1^3} \delta_{ab,cd} \delta(\omega_1, \omega_2) \delta(z_1, z_2)~,
\ee
 with $\delta_{ab,cd} = \frac{1}{2}(\delta_{ac} \delta_{bd} + \delta_{ad} \delta_{bc}) - \frac{1}{4}\delta_{ab}\delta_{cd}.$\\
One can now substitute the expansion \eqref{Dabquant} in  \eqref{softcharge} to write the quantized soft charge as\footnote{The leading and the subleading  soft mode take the following form in terms of the creation and annihilation operator
\begin{eqn}
 \intinf du \tilde D_{ab}(u, z) &= -  \frac{\sqrt{8\pi G_N}}{2(2\pi)^2} \lim_{\omega \to 0} \Big[ \tila_{ab}(\omega, z) + \tilad_{ab}(\omega, z) \Big]~, \\
 \intinf du u \tilde D_{ab}(u, z) &= \frac{i \sqrt{8\pi G_N}}{2(2\pi)^2} \lim_{\omega \to 0} (1 + \omega \p_\omega) \Big[\tila_{ab}(\omega, z) - \tilad_{ab}(\omega, z) \Big]~,
\end{eqn}
where the factor of 2 in the denominator comes from the fact that we only deal with $\omega > 0$. 
}
\begin{eqn}\label{Qsoftcharge}
Q^S_V &= \frac{i}{2\sqrt{8\pi G_N}} \int d^4 z \lim_{\omega \to 0} V^b \mathbb D^a \Big[ (1 + \omega \p_\omega) \tila_{ab} - (1 + \omega \p_\omega) \tilad_{ab}  \Big] \\
&\quad- \frac{1}{96 \pi^2 \sqrt{8\pi G_N}} \int d^4 z \lim_{\omega \to 0}  \partial^a \partial^m ({\tila}^b_{m} + {\tilde{\mrm{a}}^{\dagger b}}_m)(\lie_V \bar C_{ab} - \a \bar C_{ab})~.
\end{eqn}
Note that in order to promote the classical expression for the soft charge \eqref{softcharge} to the quantized version above, we have chosen a particular operator ordering for the terms in the second line. This choice will become clear after defining the vacuum state, as done in the following subsection. Subsequently, we will see that the Ward identity of the superrotation charges on the states built from this vacuum follows from the subleading  soft graviton theorem.

 \subsection{Vacuum state}\label{sec:vacuum}
Motivated from \cite{Campiglia:2021bap, Chowdhury:2022nus}, a convenient choice for labelling the vacua is to choose them to be the eigenstates of the operators $\bar C_{ab}$ and $T_{cd}$\footnote{Note that we use the same notation for the operators and also the classical fields.}. Soft theorems are usually studied in the Fock vacuum, which corresponds to choosing the vacuum state with  zero eigenvalue for $\bar C_{ab}$ and $T_{cd}$, i.e, $\vac$. 
States with finite energy excitations can be obtained from the vacuum state by acting with the creation operator on these states.  For example, a generic incoming state $\instate$  can be expressed as
 \be\label{instate}
 \instate =  \tilad_{h_1 \cdots h_{s_1}}(\omega_1, z_1) \cdots \tilad_{h_1 \cdots h_{s_n}}(\omega_n, z_n) \vac~,
 \ee
 where the operator $\tilad_{h_1 \cdots h_s}(\omega, z)$ denotes the creation operator for a particle of spin$-s$ with energy $\omega$ and momenta along $z$. One can similarly define the outgoing states. 
 
 For this definition, the reason for the choice of operator ordering in \eqref{Qsoftcharge} is now evident. The action of the soft charge on states defined in \eqref{instate} will not receive any contribution from the second term in \eqref{Qsoftcharge},
\begin{eqn}\label{qsc}
Q_V^S \ket{\rm in} =  \frac{i}{2\sqrt{8\pi G_N}} \intsinf d\omega V^b \mathbb D^a \Big[ (1 + \omega \p_\omega) \tila_{ab} - (1 + \omega \p_\omega) \tilad_{ab}  \Big]   \ket{\rm in}~.
\end{eqn}
There exists a similar decomposition for the hard charge but the exact structure will not be necessary for our purpose. 

 \subsection{Quantized superrotation charge for scalars coupled to gravity}
In this subsection, we consider the special case of scalar field coupled to gravity, and derive the action of the hard charge on the matter phase space. We will later use this to demonstrate how the Ward identities associated to the superrotation charges are consistent with subleading soft graviton theorem when the external states are massless scalars. 

By using the saddle point approximation, the quantized scalar field operator in the Bondi frame is given as 
\be
\phi^{(2)}(u, z) =  - \frac{1}{(2\pi)^3} \intsinf d\omega \omega \Big[ \tila(\omega, z) e^{- i u \omega} + \tilad(\omega, z) e^{i u \omega}  \Big]~,
\ee
where $\phi^{(2)}$ denotes the $\frac{1}{r^2}$ term in the large$-r$ expansion of the field $\phi(u, r, z)$, which is the dynamical mode in six dimensions. The superrotation action on $\phi^{(2)}$ is given as
\be
\delta_V \phi^{(2)} = \lim_{r\to\infty} r^2 \lie_\xi \phi = \lie_V \phi^{(2)} + u \a \p_u \phi^{(2)} + 2\a\phi^{(2)}~.
\ee
Taking an inverse Fourier transform of this equation, we obtain the spacetime action on the creation operator 
\be\label{jv}
\delta_V \tilad(\omega, z_s) = \lie_V \tilad (\omega, z_s)- \a \omega\p_\omega \tilad (\omega, z_s)= V^c \p_c \tilad (\omega, z_s)- \a \omega \p_\omega \tilad(\omega, z_s)\equiv iJ_V (\omega, z_s)~.
\ee
This is equivalent to evaluating the commutator of the hard charge with the creation operator $[Q^H_V, \tilad(\omega, z_s)]$. 

\subsection{Subleading  soft graviton theorem and Ward identity of superrotation charges}
We start by evaluating the Ward identity for the superrotation charges for massless external scalars built from the vacuum state described in section \ref{sec:vacuum}, that we expect to be implied from the subleading  soft graviton theorem. This can be written as
\begin{eqn}
\bra{\mathrm{out}}[Q_V,\mathcal{S}]\ket{\mathrm{in}} =0 \implies \bra{\mathrm{out}}[Q_V^{S},\mathcal{S}]\ket{\mathrm{in}} =-\bra{\mathrm{out}}[Q_V^{H},\mathcal{S}]\ket{\mathrm{in}}~,
\end{eqn}
with the incoming and outgoing states being massless scalars. As explained in the previous section, the charge can be written as a sum of soft and hard charge. Using the expression of soft charge \eqref{qsc} and the action of the hard charge on the external states \eqref{jv}, the Ward identity can be written as\footnote{Note that in the below equation we have used crossing symmetry to relate the incoming to outgoing subleading soft graviton mode. }
\begin{eqn}\label{mwi}
\frac{1}{ \sqrt{8\pi G_N}}\int d^4z_s\ \mathbb D^a V^b(z_s) \lim\limits_{\omega \rightarrow 0}(1 + \omega\partial_{\omega})&\bra{\mrm{out}}\tila_{ab}(\omega, z_s)\mathcal{S}\ket{\mrm{in}}\\
&\qquad= - i\Bigg(\sum_{\mrm{out}}J_V^i - \sum_{\mrm{in}}J_V^i\Bigg) \bra{\mrm{out}}\mathcal{S}\ket{\mrm{in}} ~,
\end{eqn}
where $J_V^i$ is the operator  defined in \eqref{jv} acting on the $i^{th}$ external scalar.

In six dimensions, the subleading  soft graviton theorem for the external particles being scalars can be written as 
\begin{equation}\label{ssg1}
\frac{1}{\sqrt{8\pi G_N}}\lim\limits_{\omega \rightarrow 0}(1 + \omega\partial_{\omega})\bra{\mathrm{out}}\tila_{ab}(\omega,z_s)\mathcal{S}\ket{\mathrm{in}} = -i\Bigg(\sum_{i}\frac{\epsilon_{ab}^{\mu\nu}k^i_{\nu}p^{\rho}}{p\cdot k^i }\mathcal{J}^i_{\mu\rho} \Bigg)\bra{\mathrm{out}}\mathcal{S}\ket{\mathrm{in}}~,
\end{equation}
where $\epsilon^{\mu\nu}_{ab}$ denotes the polarization tensor (with the polarization indices denoted by $a, b$) of the soft graviton with momenta $p^\mu = \omega \hat{p}^\mu$, where $\hat{p}^\mu$ denotes the unit null momenta parameterized by the flat coordinates $z_s$, $k^i$ denotes the momenta of the external scalar particle (which is parameterised by energy $\omega_{k_i}$ and $z_{k_i}$) and $\mathcal{J}^i_{\mu\nu}$ denotes the total angular momenta acting on the $i^{th}$ external particle and the sum runs over all the external particles. The subleading  soft graviton theorem \eqref{ssg1} in the flat null coordinates takes the following form \cite{Kapec:2017gsg}
\begin{eqn}\label{flsgl}
\frac{1}{\sqrt{8\pi G_N}}\lim\limits_{\omega \rightarrow 0}(1 + \omega\partial_{\omega})&\bra{\mathrm{out}}\tila_{ab}(\omega,z_s)\mathcal{S}\ket{\mathrm{in}}\\
&= \sum_{i} \Big[ P^c_{ab}(z_s - z_{k_i}) \p_{z_{k_i}^c} + \frac{1}{4} \p_c P^c_{ab}(z_s - z_{k_i}) \omega_{k_i} \p_{\omega_{k_i}} 
   \Big] \bra{\mathrm{out}}\mathcal{S}\ket{\mathrm{in}}~,
\end{eqn}
where 
\begin{equation}\label{pcab}
 P^c_{ab}(x) = \frac{1}{2} \Big( x_a \delta_b^c +  x_b \delta_a^c + \frac{1}{2} x^c \delta_{ab} - \frac{4 x_a x_b x^c}{x^2} \Big)~.
\end{equation}
We will now derive the Ward identity \eqref{mwi} from the subleading soft theorem \eqref{ssg1}. As shall be seen below, the linear terms in $P^c_{ab}(x)$ will not affect the calculation and therefore these are an artifact of the gauge choice. In order to derive the Ward identity, we smear the LHS of the soft theorem with the function $\int d^4 z_s\ \mathbb D^a V^b(z_s)$.

This will reproduce the LHS of the Ward identity \eqref{mwi}. Subsequently, by performing the same operation on the RHS of the soft theorem \eqref{flsgl}, we get the following two terms,
\begin{align}\label{slid1}
&\sum_{i} \int d^4 z_s \mathbb D^a V^b(z_s) 
P^c_{ab}(z_s - z_{k_i}) \p_{z_{k_i}^c} = \sum_{i} V^c(z_{k_i})\p_{z_{k_i}^c}~,\\\label{slid2}
&\sum_{i}\int d^4 z_s \mathbb D^a V^b(z_s)
\frac{1}{4} \p_c P^c_{ab}(z_s - z_{k_i}) \omega_{k_i} \p_{\omega_{k_i}} = \sum_{i} \a(z_{k_i}) \omega_{k_i} \p_{\omega_{k_i}}~.
\end{align}

The equations above follow from the following identity (a derivation of this result is given in appendix \ref{app:DaPcab}),
\begin{equation}\label{identity}
\frac{1}{64\pi^2} \Big[\partial^4 \partial^a P^{c}_{ab}(x)  - \frac{4}{3}\partial^2\partial^{ef}_b P^{c}_{ef}(x)\Big] = -\delta^c_b\delta^{(4)}(x)~.
\end{equation}
By taking the sum of  \eqref{slid1} and \eqref{slid2}, one recovers the $J_V$ operator in \eqref{jv} and hence, the RHS of the Ward identity \eqref{mwi} follows.

The extension of this proof to particles of arbitrary spin coupled to gravity in $d > 4$ poses some subtleties that will be addressed in a future study. 

\section{Conclusion and Discussion}\label{conclusion}
\subsection*{\it Main Result}
In this paper, we study the symmetries of non-linear general relativity in six dimensional flat spacetime far away from sources. We work with the special case where we only have massless fields and therefore it is convenient to perform such analysis near null infinity. We start by analyzing the equations of motion and the gauge conditions which enable us to  identify the free data in the theory (section \ref{flatspace}).

In section \ref{symgrav}, we find the generic set of transformations that keep the asymptotic form of the metric invariant (thereby defining fall off conditions) and also respect the gauge conditions. Such transformations are generated by two classes of vector fields, namely supertranslations and superrotations, which are the infinite dimensional extension of the Poincar\'e generators. While supertranslations leave the leading order angular metric at $\scrip$ invariant, the action of superrotation vector fields are non-trivial.
We also evaluate the action of these generators on the free data and identify the appropriate radiative mode that is consistent with the fall off conditions. 

Having found the generators of the transformations, in section \ref{SRCharges}, we compute the charges corresponding to the symmetries by demanding that they generate the correct spacetime action of the phase space variables. For simplicity, we make an assumption by restricting ourselves to variations near Bondi frames. We find that the charges split up into two pieces, one the hard piece and the other, the soft piece. Further, we find that the soft charge has a term depending on the choice of the vacuum state labeled by $\bar C_{ab}$, which is the $O(r)$ term of the metric component $g_{ab}$. Following the computations in $d = 4$ \cite{Campiglia:2021bap}, we demonstrate how the hard charge can be obtained using the covariant phase space formalism on the hard phase space. It is expected that the full symplectic structure can be derived using the Crnkovic-Witten symplectic form \cite{300years} and we leave that for future work.  
 
 Finally in section \ref{wi}, we demonstrate how the subleading soft theorems in non-linear general relativity in the Bondi frame can be used to derive the Ward identity corresponding to the superrotation charges obtained via the asymptotic symmetries.
\subsection*{\it Angular Momentum Aspect}
It was shown in earlier works \cite{Chowdhury:2022nus} that the supertranslation charge can be expressed in terms of the {\it Bondi mass aspect} $M^{(3)}$ in the non-linear theory. Motivated by the results in the linearized theory \cite{Campoleoni:2020ejn}, we similarly expect that the superrotation charge can be expressed in terms of the {\it Angular momentum aspect}, $U_a^{(3)}$, in the non-linear theory and we leave this for a future work.

\subsection*{\it Non-Bondi Frames}
In order to evaluate the charges corresponding to asymptotic symmetries, we have restricted ourselves to small variations around the Bondi frame. The deviation from the Bondi frame is controlled by the value of the tensor $T_{ab}$ and in this paper we neglect terms of $O(T^2)$. This assumption is physically motivated by the fact that in any scattering process we always set the metric on the celestial plane to be that of the metric of flat space, which is the Bondi frame ($T_{ab} = 0$). However, this also places a restriction on the classes of superrotations that are allowed and therefore a complete understanding of this would require an analysis involving an arbitrary value of $T_{ab}$. A generic analysis of this nature will also allow us to gain insight onto the asymptotic symmetries corresponding to double soft theorems \cite{Campiglia:2021bap}.

\subsection*{\it General Metric Expansions}
While defining the fall off conditions for the metric components in \eqref{bondi-1}, we have assumed that $\beta^{(0)} = 0$ and also that there are no logarithmic terms in the metric expansion. However, in \cite{Capone:2021ouo}, the author demonstrates that there exists more general fall off conditions that are compatible with 
asymptotically flat spacetimes. It will be interesting to see if the superrotation charges  receive any corrections when one includes such terms in the expansion. 


\subsection*{\it Odd dimensions}
Even though we have worked out most of the results in six dimensional spacetime, our results should be easily generalizable to any even dimension $> 4$. However, the extension of these results to odd dimensions is not straightforward (see \cite{Fuentealba:2022yqt} for some recent progress).

\section*{Acknowledgements} We are sincerely grateful to Alok Laddha for encouraging us to work on this problem, for several crucial \& insightful discussions at various stages and for his suggestions on the manuscript. We are also grateful to Ankit Aggarwal, Federico Capone, R. Loganayagam, Prahar Mitra, Ashoke Sen and all the members of ICTS Strings group for several fruitful discussions. We also thank Kasper Peeters for answering several questions related to the excellent software CADABRA which was used to evaluate some of the results \cite{Peeters2018}. Research at ICTS-TIFR is supported by the Department of Atomic Energy, Government of India, under Project Identification Nos. RTI4001. CC is grateful to IIT Indore via the ST4 workshop, where a part of this work was completed. CC and AAH would like to thank Alok Laddha and CMI for hospitality where a part of this work was completed.

\begin{appendix}
\section{Superrotation hard charge from covariant phase space}\label{app:sympform-hardcharge}
From the symplectic structure at $\scrip$ for hard phase space in non-linear general relativity \cite{Campiglia:2021bap, Chowdhury:2022nus}, we can give a proposal for the derivation of the superrotation hard charge \eqref{SRhard3} in the Bondi frame. A similar derivation for the supertranslation hard charge is already given in section 3 of \cite{Chowdhury:2022nus}. A rigorous derivation of the charge in any generic frame requires a more careful analysis of the symplectic structure at $\scrip$ which involves a study of generic variations of the background metric $q_{ab}$. This is currently under progress and we hope to address it in a future work. 

The part of the symplectic form in eq.(3.8) of \cite{Chowdhury:2022nus} contributing to the hard charge is given as 
\be\label{hardsym}
\Omega^{H}(\delta, \delta') = - \dfrac{1}{32\pi G_N} \intsp \delta \tilde D^{ab} \wedge \delta' \p_u \tilde D_{ab}~.
\ee
The superrotation hard charge is defined as
\be
\delta_0 Q_V^H = \Omega^H(\delta_0, \delta_V)~,
\ee
where the variation $\delta_0$ is defined such that $\delta_0 \delta_{ab} = 0$. Upon substituting the variations given in \eqref{Dtilvariations} we obtain\footnote{We also need the variation of the inverse $\delta_V \tilde D^{ab}$ which can be evaluated as 
\begin{eqn}
\delta_V \tilde D^{ab} &= \delta_V(q^{ac} q^{ad} \tilde D_{cd}) 
= \lie_V \tilde D^{ab} + u \a \p_u \tilde D^{ab} + 4 \a \tilde D^{ab} ~.
\end{eqn}} 
\begin{eqn}
	\delta_0 Q_V^H &=  \dfrac{1}{32\pi G_N}  \intsp \delta_0 N^{ab} \big[ 2 u \a N_{ab} + 4 \a \tilde D_{ab} \big] +  \dfrac{1}{32\pi G_N} \intsp \Big[ \delta_0 N^{ab} \lie_V \tilde D_{ab} + \delta_0 N_{ab} \lie_V \tilde D^{ab}  \Big]~,\\
	&\equiv \dfrac{1}{32\pi G_N} \big[ \delta_0 Q_{V}^{H(1)} + \delta_0 Q_{V}^{H(2)}  \big]~.
\end{eqn}
Our goal is to express the expression on RHS as a total variation in $\delta_0$. The first term in this can be simplified to give 
\be
\delta_0 Q_V^{H(1)} = \intsp u \a \delta_0(N_{ab} N^{ab}) + 4 \intsp \a \delta_0 N^{ab} \tilde D_{ab}~.
\ee
As we see below, the second term in the expression above gets canceled by a contribution arising from $\delta_0 Q_V^{H(2)}$,
\begin{eqn}
 \delta_0 Q_V^{H(2)} &= \intsp \delta_0(N^{ab} V^c \partial_c \tilde D_{ab}) + \intsp \delta_0 \big[ N^{ab} \tilde D_b^c \partial_a V_c + (a \leftrightarrow b) \big] - 4 \intsp \a \delta_0 N^{ab} \tilde D_{ab}~.
\end{eqn}

Hence upon summing up the two expressions above we get a total variation in $\delta_0$ on the RHS, which then indicates that we can perform an integration in $\delta_0$ to give $Q_V^H$ 
\begin{eqn}
Q_V^H = \dfrac{1}{32\pi G_N} \intsp N^{ab} \big[ u \a N_{ab} + \lie_V \tilde D_{ab} \big]~,
\end{eqn}
which is the same charge derived using different methods in the previous section \eqref{SRhard1}.

\section{Derivation of the identity \eqref{identity}}\label{app:DaPcab}
In\footnote{We thank R. Loganayagam for useful discussions related to this and also for suggesting several references on this topic.} this appendix we prove equation \eqref{identity}, i.e, 
\be
\frac{1}{64\pi^2} \Big[ \p^4 \p^a P^c_{ab}(x) - \frac{4}{3} \p_b \p^2 \p^{ef} P^c_{ef} \Big] = - \delta^c_b \delta^{(4)}(x)~,
\ee
where $P^c_{ab}(x) = \frac{1}{2} \big[ x_a \delta^c_b + x_b \delta^c_a + \frac{1}{2} x^c \delta_{ab} - \frac{4}{x^2} x^c x_a x_b \big] $.  We first note that the only term in $P^c_{ab}(x)$ that contributes to the equation above is the last term, i.e $\frac{1}{x^2} x^c x_a x_b$, as the terms which are linear in $x^a$ are annihilated by the derivative operators. In order to carefully handle such terms and notice the presence of delta functions, it is instructive to deform pole at $x = 0$ to a slight imaginary value by $x^2 \to x^2 + \e^2$ where the limit $\e \to 0$ has to be carefully taken in the last step of the computation, thus giving rise to terms involving delta functions in the calculation. With this in place we can evaluate the derivatives without worrying about the appearance of the delta function in the intermediate steps. Therefore we get the following terms 
\begin{eqn}
 \frac{1}{64\pi^2} \Big[  \p^4 \p^a P^c_{ab}(x) - \frac{4}{3} \p_b \p^2 \p^{ef} P^c_{ef} \Big] = \lim_{\e \to 0} \Bigg\{ \frac{1}{2\pi^2} \p_b^c \Big( \frac{\e^6}{(x^2 + \e^2)^4} \Big) - \frac{12}{\pi^2} \frac{\delta^c_b \e^6}{(x^2 + \e^2 )^5} \Bigg\}~.
\end{eqn}
In order to take the limit we need to keep in mind that these are distributions which are integrated against test functions and hence we integrate the LHS against a spherically symmetric test function $\mathcal F(|x|)$ (which has a sufficiently fast fall off) and obtain 
\begin{eqn}
&\int d^4 x \mathcal F(|x|)  \frac{1}{64\pi^2} \Big[  \p^4 \p^a P^c_{ab}(x) - \frac{4}{3} \p_b \p^2 \p^{ef} P^c_{ef} \Big] \\
&= \lim_{\e \to 0} \int d^4 x \mathcal F(|x|)  \Bigg\{ \frac{1}{2\pi^2} \p_b^c \left[ \frac{\e^6}{(x^2 + \e^2)^4} \right] - \frac{12}{\pi^2} \frac{\delta^c_b \e^6}{(x^2 + \e^2 )^5} \Bigg\}~,\\
&=   \lim_{\e \to 0} \int_0^\infty d |x| |x|^3 |\mathbb S_3|\mathcal F(|x|)  \Bigg\{ \frac{1}{2\pi^2} \p_b^c \left[ \frac{\e^6}{(|x|^2 + \e^2)^4} \right] - \frac{12}{\pi^2} \frac{\delta^c_b \e^6}{(|x|^2 + \e^2 )^5} \Bigg\} ~.
\end{eqn}
where $\mathbb S_3 = 2\pi^2$. As shown in \cite{1983AmJPh51826F}, such expressions are simplified by expanding $\mathcal F(x)$ in a Taylor series and upon performing the integrals and taking the limit $\e \to 0$ we obtain 
\be
\int d^4 x \mathcal F(|x|)  \frac{1}{64\pi^2} \Big[  \p^4 \p^a P^c_{ab}(x) - \frac{4}{3} \p_b \p^2 \p^{ef} P^c_{ef} \Big] = - \delta_b^c \mathcal F(0)~.
\ee
As this is true for any generic test function $\mathcal F(x)$  we conclude that
\begin{eqn*}
\frac{1}{64\pi^2} \Big[ \p^4 \p^a P^c_{ab}(x) - \frac{4}{3} \p_b \p^2 \p^{ef} P^c_{ef} \Big] = - \delta^c_b \delta^{(4)}(x)~,
 \end{eqn*}
thereby proving the identity \eqref{identity}.

\end{appendix}
\printbibliography 
\end{document}